\documentclass[submission,copyright,creativecommons]{eptcs}
\providecommand{\event}{ICE 2015} 
\usepackage{graphicx}
\DeclareGraphicsExtensions{.pdf,.png,.jpg}
\usepackage{makeidx}
\usepackage{epsf}
\usepackage{latexsym}
\usepackage{amsfonts}
\usepackage[fleqn]{amsmath}
\usepackage{amssymb}
\usepackage{url}
\usepackage{scalefnt}
\usepackage{rotating}
\usepackage{xcolor} 
\usepackage{listings}
\definecolor{darkgray}{gray}{0.25}
\lstdefinelanguage{APT}{
        alsoletter={.},
        morekeywords={.name, .description, .type, .states, .initial, .labels, .arcs, .places, .transitions, .flows, .initial_marking},
        otherkeywords={->},
        comment=[s]{/*}{*/},
        morecomment=[l]{//},
        commentstyle=\color{darkgray}\ttfamily,
        morestring=[b]"
}
\lstdefinelanguage{shell}{
}
\usepackage{tikz}
\usetikzlibrary{decorations,decorations.pathreplacing,decorations.pathmorphing,decorations.markings}
\usetikzlibrary{arrows}
\usetikzlibrary{shapes.geometric}
\usetikzlibrary{shapes.symbols}

\makeindex

\begin{document}

\newcommand{\p}[1]{\mbox{{#1}\hspace{-0.20em}\c{}\hspace{0.20em}}}
\newcommand{\anz}{{\tt \#}}\newcommand{\infunnel}{{}^\circ}
\newcommand{\es}{\emptyset}
\newcommand{\pminus}{
\mbox{\textrm{$-$}\!\!\!\!\!\!\!\:\,\,\raisebox{1.5mm}{$\scriptstyle \bullet$}}\:}
\newcommand{\fto}[1]{\stackrel{#1}{\rightarrow}}
\newcommand{\ftov}[1]{\downarrow\!{\scriptstyle #1}}
\newcommand{\dt}{\bullet}
\newcommand{\leer}{\varepsilon}
\newcommand{\nsymbol}{\mathbb{N}}
\newcommand{\zsymbol}{\mathbb{Z}}
\newcommand{\qsymbol}{\mathbb{Q}}
\newcommand{\minus}{\setminus}
\newcommand{\goesto}{\rightarrow}
\renewcommand{\goesto}[1]{\stackrel{#1}{\longrightarrow}}
\newcommand{\longgoesto}[1]{\stackrel{#1}{-\!\!\!-\!\!\!-\!\!\!-\!\!\!-\!\!\!-\!\!\!-\!\!\!-\!\!\!-\!\!\!-\!\!\!\!\!\!\longrightarrow}}
\newcommand{\verylonggoesto}[1]{\stackrel{#1}{-\!\!\!-\!\!\!-\!\!\!-\!\!\!-\!\!\!-\!\!\!-\!\!\!-\!\!\!-\!\!\!-\!\!\!-\!\!\!-\!\!\!-\!\!\!-\!\!\!-\!\!\!-\!\!\!-\!\!\!-\!\!\!-\!\!\!-\!\!\!-\!\!\!-\!\!\!-\!\!\!\!\!\!\longrightarrow}}
\newcommand{\ggoesto}[1]{\stackrel{#1}{\Longrightarrow}}
\newcommand{\emptyseq}{\varepsilon}
\newcommand{\impl}{\Rightarrow}
\newcommand{\Parikh}{\Psi}
\def\tp{^{\sf T}}
\newcommand{\tile}[4]{\begin{array}{|ll|}\hline#1&#2\\#3&#4\\\hline\end{array}}
\newcommand{\disjcup}{\makebox[1.0em][c]{
\setlength{\unitlength}{0.4em}
\begin{picture}(2,2)(0,0)
\put(0.90,0.4){{\scriptsize $\bullet$}}
\put(0.5,0){$\cup$}
\end{picture}
\setlength{\unitlength}{1mm}
}}

\newcommand{\EndSy}{\hfill\protect\makebox[1.0em][c]{
\protect\setlength{\unitlength}{0.2em}
\protect\begin{picture}(3,3)(0,0)
        \begin{thinlines}
\protect\put(0,0){\line(1,0){3}}
\protect\put(0,0){\line(0,1){3}}
\protect\put(0,3){\line(1,0){3}}
\protect\put(3,0){\line(0,1){3}}
\protect\put(0,0){\line(1,1){3}}
\protect\put(0,3){\line(1,-1){3}}
        \end{thinlines}
\protect\end{picture}
\protect\setlength{\unitlength}{1mm}
}}

\newcommand{\choice}{\protect\makebox[1.0em][c]{
\protect\setlength{\unitlength}{0.2em}
\protect\begin{picture}(2,4)(0,0)
\protect\put(0,0){\line(1,0){2}}
\protect\put(0,0){\line(0,0){4}}
\protect\put(0,4){\line(1,0){2}}
\protect\put(2,0){\line(0,1){4}}
\protect\end{picture}
\protect\setlength{\unitlength}{1mm}
}}

\newcommand{\BX}[1]{{\unskip\nobreak\hfil\penalty50
                    \hskip2em\hbox{}\hfil
\EndSy\/ {{\rm #1}}
                    \parfillskip=0pt \finalhyphendemerits=0 \par
                   }}

\newcommand{\DEF}[2]{\goodbreak\begin{definition}
                     \label{#1}\begin{rm}{\sc #2}

                    }
\newcommand{\ENDDEF}[1]{\BX{\ref{#1}}
                        \end{rm}\end{definition}
                       }
\newcommand{\ENXDEF}{\end{rm}\end{definition}}
\newcommand{\KRYPT}[2]{\goodbreak\begin{kryptosystem}
                     \label{#1}\begin{rm}{\sc #2}

                    }
\newcommand{\ENDKRYPT}[1]{\BX{\ref{#1}}
                        \end{rm}\end{kryptosystem}
                       }
\newcommand{\KOR}[2]{\goodbreak\begin{corollary}
                     \label{#1}{\sc #2}

                    }
\newcommand{\ENDKOR}[1]{\BX{\ref{#1}}
\end{corollary}
                       }
\newcommand{\ENXKOR}{
\end{corollary}}
\newcommand{\PROP}[2]{\goodbreak\begin{proposition}
                      \label{#1}{\sc #2}

                     }
\newcommand{\ENDPROP}{
\end{proposition}}
\newcommand{\ENXPROP}[1]{\BX{\ref{#1}}
\end{proposition}
                       }
\newcommand{\THEO}[2]{\goodbreak\begin{theorem}
                     \label{#1}{\sc #2}

                    }
\newcommand{\ENDTHEO}{
\end{theorem}}
\newcommand{\SATZ}[2]{\goodbreak\begin{theorem}
                     \label{#1}{\sc #2}

                    }
\newcommand{\ENDSATZ}{
\end{theorem}}
\newcommand{\ENXSATZ}[1]{\BX{\ref{#1}}
\end{theorem}}
\newcommand{\LEM}[2]{\goodbreak\begin{lemma}
                     \label{#1}{\sc #2}

                    }
\newcommand{\ENDLEM}{
\end{lemma}}
\newcommand{\ENXLEM}[1]{\BX{\ref{#1}}
\end{lemma}}
\newcommand{\BEW}{\goodbreak
{\bf Proof:}
                 }
\newcommand{\XBEW}{\goodbreak{\bf Proof:} }
\newcommand{\ENDBEW}[1]{\BX{\ref{#1}}
                       }
\newcommand{\ENXBEW}{}
\newcommand{\BBEW}[1]{\goodbreak{\bf Beweis von {\rm #1}:}}

\newcommand{\ENDBBEW}[1]{\BX{\ref{#1}}
                        }
\newcommand{\ENXBBEW}{}
\newcommand{\NOT}[2]{\goodbreak\begin{notation}
                     \label{#1}\begin{rm}{\sc #2}

                    }
\newcommand{\ENDNOT}[1]{\BX{\ref{#1}}
                        \end{rm}\end{notation}
                       }
\newcommand{\ENXNOT}{\end{rm}\end{notation}}
\newcommand{\REM}[2]{\goodbreak\begin{remark}
                     \label{#1}\begin{rm}{\sc #2}
                    }
\newcommand{\ENDREM}[1]{\BX{\ref{#1}}
                        \end{rm}\end{remark}
                       }
\newcommand{\ENXREM}{\end{rm}\end{remark}}
\newcommand{\BSP}[2]{\goodbreak\begin{beispiel}
                     \label{#1}\begin{rm}{\sc #2}

                    }
\newcommand{\ENDBSP}[1]{\BX{\ref{#1}}
                        \end{rm}\end{bespiel}
                       }

\newcommand{\prefix}{\sqsubseteq}
\newcommand{\backwd}[1]{\stackrel{\leftharpoonup}{#1}}
\newcommand{\down}{\;\downarrow\!}
\renewcommand{\vec}[1]{\overrightarrow{#1}}
\newcommand{\uniqueP}{\Upsilon}
\newcommand{\vzero}{0}
\newcommand{\drop}[1]{}

\renewcommand{\labelitemi}{$\bullet$}
\renewcommand{\labelitemii}{$\circ$}
\renewcommand{\labelitemiii}{$\cdot$}
\renewcommand{\labelitemiv}{$\ast$}
\itemsep0pt

\title{Analysis of Petri Nets and Transition Systems}
\author{Eike Best and Uli Schlachter%
\thanks{The authors are supported by the German Research Foundation (DFG)
project ARS (Algorithms for Reengineering and Synthesis), reference number Be 1267/15-1.} 
\institute{Department of Computing Science, Carl von Ossietzky Universit\"at \\
D-26111 Oldenburg, Germany}
\email{\{eike.best,uli.schlachter\}@informatik.uni-oldenburg.de}
}
\def\titlerunning{Analysis of Petri Nets and Transition Systems}
\def\authorrunning{Eike Best \& Uli Schlachter}


\maketitle

\begin{abstract}
This paper describes a stand-alone, no-frills tool supporting the
analysis of (labelled) place/transition Petri nets
and the synthesis of labelled transition systems into Petri nets.
It is implemented as a collection of independent, dedicated algorithms
which have been designed to operate modularly, portably, extensibly, and efficiently.
\end{abstract}


\section{Motivation}
\label{motiv.sct}

Labelled transition systems are frequently employed in order
to display the state space of a given Petri net
and to analyse its behavioural properties.
Conversely, by region theory \cite{bbd}, a Petri net may be synthesisable
from a given labelled transition system.
Such a net is then correct ``by design''.
However, a transition system may be extremely (even infinitely) large,
causing synthesis algorithms to be prohibitively time-consuming.
Moreover, synthesis suffers from nondeterminism, since for a given transition system,
many different Petri net implementations may exist.

In such a context, it is interesting to discover relationships between
special, albeit useful, classes of transition systems and classes
of Petri nets (e.g., persistent ones \cite{tcs97}), so that faster and more deterministic
analysis and synthesis methods can be devised.
For the working mathematician, this tends to involve the error-prone examination of graphs
which may be large and intricate.
Tools such as {\tt synet} \cite{synet} and {\tt petrify} \cite{petrify} are helpful, 
but there is also a need for multifunctional tools with the following properties:
\begin{itemize}
\item
Versatility.
The user should be able to create, modify, and manage hundreds or thousands of medium-sized graphs
(both Petri nets and transition systems) which might only slightly be at variance with each
other. 
E.g., in {\tt synet}, the only way of inserting comments on data objects is by choosing meaningful file names.
For large collections of objects, a more flexible commenting function becomes mandatory.
No restrictions should be imported from intended applications.
E.g., {\tt petrify} excludes non-safe Petri nets as output because
they are of no interest in a hardware context.
\item
Transparency.
The tool's internal machinations should be detectable, if necessary by examining the source code.
E.g.,
it is not known whether {\tt synet} always constructs a safe Petri net if there exists one.
\item
Extensibility.
It should be possible to program and add modules fast, in case
the need arises for any particular new problems.
In particular, modules should have properly defined, readable, and descriptive input/output interfaces.
\item
Bare-bonedness.
The tool should operate on place/transition nets with arbitrary arc weights and side-conditions,
and on arbitrary labelled transition systems, as well as on many interesting sub- (rather than super-) classes.
Emphasis should be on algorithmic optimisation, rather than on textual expressiveness.
Communication between users, as well as between tools, should be achieved via human-readable text files.
\item
Efficiency and modularity.
Analysis of medium-sized objects (say, graphs with a few hundred nodes) should
be fast, even if the theoretical complexity is {\tt PSPACE}-hardness or worse.
In the event of bottlenecks, the tool should be sufficiently modular
so that the culprit(s) can be isolated quickly.
Memory should be organised in such a way that average-sized objects can be handled
and overflow does not occur, or can at least be localised cleanly.
\item
Portability and availability.
It should be possible to switch quickly between different platforms.
No frequent recompiling should occur,
and any dependencies on residual installations should be minimised.
The tool should be freely downloadable and usable as a single executable file on many different platforms.
No registration or other ``paperwork'' (such as sending mails or waiting for release links),
and few system-dependent installations, should be necessary in order to use it.
\end{itemize}
Since a tool of this kind was found to be lacking, 
a students' project was initiated at the University of Oldenburg in 2012.
The toolbox that resulted from it by March 2013 has been called {\tt APT}
for {\em {\tt A}nalysis of {\tt P}etri nets and {\tt T}ransition systems}
and is available at \cite{apt}.
Since then, {\tt APT} has been optimised and extended by the second author (and other persons).
The present paper contains a brief summary of the use and structure of {\tt APT}
in sections \ref{first-steps.sct} and \ref{apt.sct}, respectively.
Some recent developments will be described in section \ref{synthesis.sct}.
Formal definitions can be found in section \ref{background.sct}.
Many of them conform with \cite{BD-Tunis14,best-wimmel-rostock} where a more detailed exposition of some of the theory can be found.

\section{Introduction to the use of {\tt APT}, and some examples} 
\label{first-steps.sct}

\newcommand{\apt}[0]{\texttt{APT}}

\apt{} is implemented in Java 7 and is released under the \texttt{GPLv2} license.
As one of the goals was portability, it consists of a single file called \texttt{apt.jar} 
which can be run by any Java 7 runtime environment.
Currently there is no graphical user interface, but instead a console-based one.
This decision was made to be able to focus on the implementation of algorithms.
Listing \ref{lst:building} shows how \apt{} can be downloaded with \texttt{git} and built with \texttt{ant}.
As an alternative to using \texttt{ant}, the file {\tt apt.jar} can simply be copied from another machine.
Presently, no pre-compiled versions are available for download.
Listing \ref{lst:building} also illustrates the use of \apt{}'s \texttt{help} function.

\begin{lstlisting}[float, language=shell, label={lst:building}, caption={Downloading and building \apt{}.
Some output is omitted for reasons of brevity.}]
$ git clone http://github.com/CvO-theory/apt.git
$ cd apt
$ ant jar
$ java -jar apt.jar help bounded
Usage: apt bounded <pn> [<k>]
  pn         The Petri net that should be examined
  k          If given, k-boundedness is checked
Check if a Petri net is bounded or k-bounded.
\end{lstlisting}

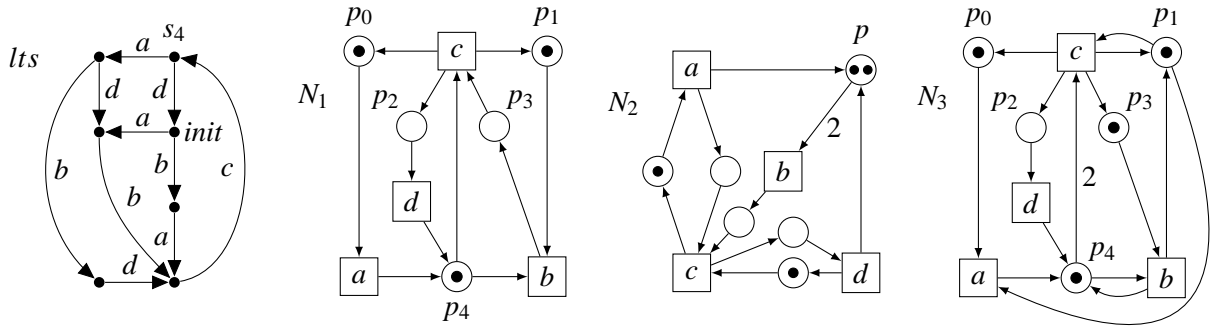
\begin{figure}[htb]
\centering
\raisebox{0.3cm}{\begin{tikzpicture}[scale=1.0]
\node[circle,fill=black!100,inner sep=0.05cm](s0)at(2,2)[label=above right:]{};\node[right of=s0,node distance=11pt]{$\mathit{init}$};
\node[circle,fill=black!100,inner sep=0.05cm](s1)at(1,2){};
\node[circle,fill=black!100,inner sep=0.05cm](s2)at(2,1)[label=above right:]{};\node[right of=s2,node distance=11pt]{};
\node[circle,fill=black!100,inner sep=0.05cm](s3)at(2,0)[label=below:]{};
\node[circle,fill=black!100,inner sep=0.05cm](s4)at(2,3)[label=above:$s_4$]{};
\node[circle,fill=black!100,inner sep=0.05cm](s5)at(1,0)[label=below:]{};
\node[circle,fill=black!100,inner sep=0.05cm](s6)at(1,3){};
\draw[-triangle 45](s0)--node[auto,swap,inner sep=2pt,pos=0.4]{$a$}(s1);
\draw[-triangle 45](s0)--node[auto,swap,inner sep=2pt,pos=0.4]{$b$}(s2);
\draw[-triangle 45](s1)to[out=270,in=135]node[auto,inner sep=3pt]{$b$}(s3);
\draw[-triangle 45](s2)--node[auto,swap,inner sep=2pt,pos=0.4]{$a$}(s3);
\draw[-triangle 45](s3)to[out=10,in=-10]node[auto,inner sep=4pt]{$c$}(s4);
\draw[-triangle 45](s4)--node[auto,swap,inner sep=2pt,pos=0.4]{$a$}(s6);
\draw[-triangle 45](s6)to[out=220,in=140]node[auto,inner sep=3pt]{$b$}(s5);
\draw[-triangle 45](s5)--node[auto,inner sep=3pt,pos=0.4]{$d$}(s3);
\draw[-triangle 45](s4)--node[auto,swap,inner sep=2pt,pos=0.4]{$d$}(s0);
\draw[-triangle 45](s6)--node[auto,inner sep=2pt,pos=0.4]{$d$}(s1);
\node[]at(0,3){$lts$};
\end{tikzpicture}}\hspace*{0.3cm}\begin{tikzpicture}
\node[draw,minimum size=0.5cm](a)at(0.2,0){$a$};
\node[draw,minimum size=0.5cm](b)at(2.7,0){$b$};
\node[draw,minimum size=0.5cm](c)at(1.5,3){$c$};
\node[draw,minimum size=0.5cm](d)at(0.9,1){$d$};
\node[circle,draw,minimum size=0.4cm](s0)at(0.2,3)[label=above:$p_0$]{};\filldraw[black](0.2,3)circle(2pt);
\node[circle,draw,minimum size=0.4cm](s1)at(0.9,2){};\node[above left of=s1,node distance=14pt]{$p_2$};
\node[circle,draw,minimum size=0.4cm](s2)at(2,2)[label=above right:]{};\node[above right of=s2,node distance=14pt]{$p_3$};
\node[circle,draw,minimum size=0.4cm](s3)at(2.7,3)[label=above:$p_1$]{};\filldraw[black](2.7,3)circle(2pt);
\node[circle,draw,minimum size=0.4cm](s4)at(1.5,0)[label=below:$p_4$]{};\filldraw[black](1.5,0)circle(2pt);
\draw[-latex](c)--(s0);\draw[-latex](s0)--(a);
\draw[-latex](c)--(s1);\draw[-latex](s1)--(d);
\draw[-latex](b)--(s2);\draw[-latex](s2)--(c);
\draw[-latex](c)--(s3);\draw[-latex](s3)--(b);
\draw[-latex](a)--(s4);\draw[-latex](d)--(s4);
\draw[-latex](s4)--(c);\draw[-latex](s4)--(b);
\node[]at(-0.4,2.4){$N_1$};
\end{tikzpicture}\hspace*{0.3cm}
\raisebox{0.5cm}{\begin{tikzpicture}[scale=0.9]
\node[draw,minimum size=0.5cm](a)at(0,3){$a$};
\node[draw,minimum size=0.5cm](b)at(1.35,1.5){$b$};
\node[draw,minimum size=0.5cm](c)at(0,0){$c$};
\node[draw,minimum size=0.5cm](d)at(2.5,0){$d$};
\node[circle,draw,minimum size=0.4cm](p1)at(-.5,1.5)[label=right:]{};\filldraw[black](-.5,1.5)circle(2pt);
\node[circle,draw,minimum size=0.4cm](p2)at(.5,1.5)[label=above right:]{};
\node[circle,draw,minimum size=0.4cm](p3)at(2.5,3)[label=above:$p$]{};\filldraw[black](2.4,3)circle(1.8pt);\filldraw[black](2.6,3)circle(1.8pt);
\node[circle,draw,minimum size=0.4cm](p4)at(0.7,0.75)[label=right:]{};
\node[circle,draw,minimum size=0.4cm](p5)at(1.5,.6)[label=above right:]{};
\node[circle,draw,minimum size=0.4cm](p6)at(1.5,0)[label=above:]{};\filldraw[black](1.5,0)circle(2pt);
\draw[-latex](p1)--(a);\draw[-latex](a)--(p2);\draw[-latex](a)--(p3);
\draw[-latex](c)--(p1);\draw[-latex](p2)--(c);\draw[-latex](p4)--(c);\draw[-latex](c)--(p5);\draw[-latex](p6)--(c);
\draw[-latex](b)--(p4);\draw[-latex](p3)--(b);\draw(2.1,2.15)node[]{$2$};
\draw[-latex](d)--(p6);\draw[-latex](d)--(p3);\draw[-latex](p5)--(d);
\node[]at(-1,2.5){$N_2$};
\end{tikzpicture}}\hspace*{0.3cm}
\raisebox{-0.0cm}{\begin{tikzpicture}[scale=1.0]
\node[]at(-0.4,2.4){$N_3$};
\node[draw,minimum size=0.5cm](a)at(0.2,0){$a$};
\node[draw,minimum size=0.5cm](b)at(2.7,0){$b$};
\node[draw,minimum size=0.5cm](c)at(1.5,3){$c$};
\node[draw,minimum size=0.5cm](d)at(0.9,1){$d$};
\node[circle,draw,minimum size=0.4cm](s0)at(0.2,3)[label=above:$p_0$]{};\filldraw[black](0.2,3)circle(2pt);
\node[circle,draw,minimum size=0.4cm](s1)at(0.9,2){};\node[above left of=s1,node distance=14pt]{$p_2$};
\node[circle,draw,minimum size=0.4cm](s2)at(2,2)[label=above right:]{};\node[above right of=s2,node distance=14pt]{$p_3$};\filldraw[black](2,2)circle(2pt);
\node[circle,draw,minimum size=0.4cm](s3)at(2.7,3)[label=above:$p_1$]{};\filldraw[black](2.7,3)circle(2pt);
\node[circle,draw,minimum size=0.4cm](s4)at(1.5,0)[label=below:]{};\node[above right of=s4,node distance=14pt]{$p_4$};\filldraw[black](1.5,0)circle(2pt);
\draw[-latex](c)--(s0);\draw[-latex](s0)--(a);
\draw[-latex](c)--(s1);\draw[-latex](s1)--(d);
\draw[-latex](c)--(s2);\draw[-latex](s2)--(b);
\draw[-latex](s3)edge [bend right](c);
\draw[-latex](a)--(s4);\draw[-latex](d)--(s4);
\draw[-latex](s4)--(b);
\draw[-latex](s4)--node[auto,swap,inner sep=2pt,pos=0.45]{$2$}(c);
\draw[-latex](b)edge [bend left](s4);
\draw[-latex](c)--(s3);
\draw[-latex](b)--(s3);
\clip (0, -0.7) rectangle (3.3, 3.5); 
\draw[-latex](s3) .. controls (3.7,0.3) and (3.5,-1.5) .. (a);
\end{tikzpicture}}
\caption{A persistent, reversible $lts$ having the strong small cycle property with Parikh vector $1$. \protect\\
Three Petri nets $N_1,N_2,N_3$ solving it are also shown. The $lts$ has no marked graph solution.}
\label{no-mg.fig}\label{fig:example_PN}
\end{figure}

\begin{lstlisting}[float, language=APT, label={lst:example_PN}, caption={File {\tt net.apt} containing $N_1$, as depicted in figure \ref{fig:example_PN},
in \apt{} text file format.}]
.name "file name: net.apt; file content: a persistent and reversible net"
.description "A Petri net N_1 having the small cycle property"
.type LPN /* stands for Labelled Petri Net */
.places
p0 p1 p2 p3 p4 /* five places */
.transitions
a b c d /* four transitions */
.flows
a: { p0 } -> { p4 }
b: { p4, p1 } -> { p3 }
c: { p4, p3 } -> { p0, p1, p2 }
d: { 1 * p2 } -> { 1 * p4 }         // 1 * is actually redundant
.initial_marking { p0, 1 * p1, p4 } // same here
\end{lstlisting}

Figure \ref{fig:example_PN} shows a labelled transition system, $lts$, and three Petri nets, $N_1$--$N_3$,
serving as running examples.
All three Petri nets are solutions of $lts$, that is, their reachability graphs are isomorphic to $lts$.
Listing \ref{lst:example_PN} represents $N_1$ in {\tt APT}'s file format.
The file starts with a name and a description of the net.
$N_1$ has five places named {\tt p0} to {\tt p4},
and four transitions, {\tt a} to {\tt d}.
The flows of the net are specified in multiset notation.
For example, transition {\tt a} takes a token from place {\tt p0} and puts it on {\tt p4}.
Weights can be specified either by mentioning a place multiple times, e.g. {\tt \{p,p\}},
or by explicitly specifying a weight, as in
{\tt {2*p}}.
The initial marking of the net is represented in a similar format.
Comments can be enclosed within {\tt /*..*/} or begin with {\tt //} and extend to the end of the line.
{\tt APT}'s transition-centred way of specifying place/transition nets allows
multiset arc weights and markings to be represented readably.
For switching quickly between {\tt APT} and {\tt synet} formats,
{\tt APT} contains two translation modules {\tt synet2apt} and {\tt apt2synet}.
Third-party formats for Petri nets, such as the {\tt LoLA} \cite{lola} and {\tt PNML} (cf. \url{http://www.pnml.org/}) formats,
are also supported.

\begin{lstlisting}[float, language=shell, label={lst:bounded}, caption={Illustration of how to use the \texttt{bounded} module. \protect\\
On Unix-like platforms, the shell script \texttt{apt.sh}
serves as a shorthand for starting \apt{}.}]
$ ./apt.sh bounded net.apt
bounded: Yes
$ ./apt.sh bounded net.apt 1
bounded: No
witness_place: p4
witness_firing_sequence: [a]
\end{lstlisting}

The \apt{} toolbox provides a large number of \emph{modules}.
If the program is started without any arguments, a full list of available modules is printed.
A special module called \texttt{help}
(already illustrated in listing \ref{lst:building} for the module \texttt{bounded})
can be used for obtaining information about a module.
It can be seen that the \texttt{bounded} module requires a Petri net as input and optionally
accepts a value {\tt k} to check for \(k\)-boundedness.
In listing \ref{lst:bounded} both features are exemplified.
The results show that $N_1$ (of figure \ref{fig:example_PN}) is bounded, but not 1-bounded.
For 1-boundedness, \apt{} provides a witness for the negative result, stating that
after firing transition {\tt a}, place {\tt p4} will have more than one token on it.

\begin{lstlisting}[float, language=APT, label={lst:example_LTS}, caption={Reachability graph of $N_1$,
generated with \texttt{./apt.sh coverab net.apt lts.apt} and slightly edited, in order to minimise the number of lines.
It is isomorphic to $lts$ shown in figure \ref{fig:example_PN}.}]
.name ""   /* file name: lts.apt (this comment was added manually) */
.type LTS  /* stands for Labelled Transition System (this comment was added manually) */
.states
s0[initial] /* [ [p0:1] [p1:1] [p2:0] [p3:0] [p4:1] ] */
s1          /* [ [p0:0] [p1:1] [p2:0] [p3:0] [p4:2] ] */
s2          /* [ [p0:1] [p1:0] [p2:0] [p3:1] [p4:0] ] */
s3          /* [ [p0:0] [p1:0] [p2:0] [p3:1] [p4:1] ] */
s4          /* [ [p0:1] [p1:1] [p2:1] [p3:0] [p4:0] ] */
s5          /* [ [p0:0] [p1:1] [p2:1] [p3:0] [p4:1] ] */
s6          /* [ [p0:0] [p1:0] [p2:1] [p3:1] [p4:0] ] */
.labels
a b c d
.arcs
s0 a s1     s0 b s2     s1 b s3     s2 a s3     s3 c s4
s4 a s5     s4 d s0     s5 b s6     s5 d s1     s6 d s3
\end{lstlisting}

The \texttt{coverability\_graph} module of \apt{} can be used to generate a coverability graph \cite{best-wimmel-rostock} of a Petri net.
For a bounded net, this will be the reachability graph (cf. section \ref{background.sct}).
Listing \ref{lst:example_LTS} shows the reachability graph calculated by
\apt{} for our running example via \texttt{./apt.sh coverability\_graph net.apt}. Module names can be
shortened, as long as the resulting prefix is unique. So we can also use \texttt{coverab} to call the coverability module.
The initial state is always called {\tt s0}.
The correspondence between states and markings is given as a comment.
The \texttt{draw} module can be used to translate the calculated graph into the DOT format used by the {\tt GraphViz} tool
(cf. \url{http://www.graphviz.org/}) which can then visualize the graph.

Note that, in $lts$, each small cycle contains every transition exactly once.
Such a property can be examined with \apt{}.
The module \texttt{compute\_pvs}
can be used to compute the Parikh vectors of all small cycles of an lts,
and the module \texttt{cycles\_same\_pv}
checks whether all small cycles have the same Parikh vector.

\section{Overview of {\tt APT}}
\label{apt.sct}

Four stages can be distinguished in the development of {\tt APT}: 
an implementation of the necessary data structures, 
various analysis modules, 
and Petri net creator modules, described in this section, 
as well as, more recently, an implementation of Petri net synthesis, described in section \ref{stage4.sct} below.

{\bf Data structures of {\tt APT}.}
At the heart of the {\tt APT} toolbox sits a module system that ensures a high level
of extensibility and modularity.
Every module consist of an input specification, an output specification, and an
algorithm.
After a module has been registered with the module system, it is automatically
available to be used from the command line.
It is possible to create new modules by using the \texttt{Module} interface.
The methods of this interface are responsible for the definition of the
algorithm and the specification of parameters and return values, including their
names, descriptions and types. This also includes a free text description that can
include, for example, formal definitions
and usage samples.
Algorithms are implemented by the \texttt{run} method.
Within this method, an algorithm can access the parameters that were entered
by the user on the command-line.
These parameters are automatically transformed
into Java objects with the expected types according to the input specifications.
The transformations from the textual representation to Java objects and {\it vice versa}
happens automatically, and thus,
a module can focus on working with the actual objects such as
Petri nets or labelled transition systems, without needing to worry about user input / output.

For the underlying data structures implementing the objects LPN and LTS, no existing library was used, but instead, 
inspiration was drawn from the Petri Net API (\url{http://service-technology.org/pnapi/})
to design robust and versatile data structures.
The main idea is the central management of data. The
\texttt{PetriNet} class, respectively the \texttt{TransitionSystem} class, is used
as a factory to create or delete nodes, arcs, etc. Every modification of the
graph has to be done from the graph class itself, or is forwarded to it.
For data storage, a compromise between memory and running time has been made.
For example, the pre- and postsets of all nodes are stored by means of
Java's \texttt{SoftReference}s. Hence, as long as enough
memory is available, the pre- and postsets of all nodes are saved to gain a fast
access to the sets.
Otherwise the garbage collector of Java's Virtual Machine is
allowed to delete as many pre- and postsets as necessary to achieve free memory.
In this case the pre- and postsets are re-calculated and re-saved, once they
are needed.

{\bf Some stand-alone analysis modules.}
In each case, a (negative) answer is accompanied by (counter-) examples as appropriate.
The list can be extended as the need arises.

For a given finite lts (with initial state $s_0$), 
\begin{itemize}
\item
Check determinism, total reachability, persistence, reversibility,
and the small cycle property; 
\item
Compute weakly / strongly connected components and Parikh vectors of small cycles; 
\item
Check (distributed) Petri net generability by 
two external programs, {\tt synet} \cite{synet} and {\tt petrify} \cite{petrify}.
\end{itemize}
For a given Petri net (with initial marking $M_0$), 
\begin{itemize}
\item
Check the existence of isolated elements, plainness, pureness,
the existence of non-plain side conditions, weak / strong connectedness, coveredness by S-invariants / T-invariants,
the marked graph / T-net / ON / CF / other structural properties, the BCF / BiCF properties,
($k$-) boundedness, (weak) liveness, persistence, reversibility,
the small cycle properties as with lts, and weak / strong separability; 
\item
Compute all connected components, the backward, forward, and incidence matrices,
all side conditions, all (minimal, semipositive) S- and T-invariants, all minimal siphons / traps,
the greatest common divisor of the initial marking,
and {\bf if} bounded {\bf then} reachability graph {\bf else} coverability graph {\bf f{}i}.\footnote{
Several
of the other tasks require boundedness as a precondition,
so that the boundedness check is often used as a first step.}
\end{itemize}
For a given labelled Petri net with initial marking $M_0$ and labelling $h\colon T\to\Sigma$, 
\begin{itemize}
\item
Check whether a given word $w\in\Sigma^*$ is in the language of the net,
check language equivalence, and check isomorphism and bisimulation of reachability graphs.
\end{itemize}
The tasks described in this list are obviously of
very diverse degrees of complexity.
One amongst them ({\em Given a Petri net, is it separable?})
has an unknown decidability status.
Therefore, a restrictive algorithm was implemented in this module,
allowing bounds to be specified for the lengths of firing sequences.
%

{\bf Generator modules in {\tt APT}.}
These modules are useful, e.g., for benchmarking purposes (cf. section \ref{benchmark.sct}). 
\begin{itemize}
\item
Generate regular sample nets, for instance:
$n$-bit marked graph nets, for some specification or range of $n$;
$n$-philosopher nets \cite{dij-phil};
all marked graphs with a limited number of places, transitions, and tokens.
\end{itemize}
{\bf Counterexample finding modules.}
These modules (understandably) suffer from runtime problems. 
\begin{itemize}
\item
For a net, check whether the preconditions of the
conjecture mentioned at the end of section \ref{background.sct} are satisfied,
and then check isomorphism against all marked graphs of a limited size.
Do the same for a small number of randomly selected marked graphs of bigger sizes. \\
\item
Try to find intelligent extensions of an lts, such that
the preconditions of the same conjecture remain satisfied.
Find minimal extensions of an lts that satisfy all required properties.
\end{itemize}

\section{Petri net synthesis with \apt{}}
\label{stage4.sct}\label{synthesis.sct}

The goal of net synthesis is to find an injectively labelled Petri net whose reachability graph is isomorphic to a given lts.
\apt's \texttt{synthesize} module (a recent addition to \apt{} by the second author)
accepts up to three parameters.
The second parameter is the transition system from which a Petri net should be synthesised
and the third parameter can optionally specify where the calculated Petri net could be saved.
The first parameter is a comma-separated list of properties that the produced
Petri net should satisfy.
Supported properties are, at present: 
\texttt{none}, which can be used if just a generic P/T net without special properties is needed; 
\texttt{pure} to synthesise a net without side-conditions; 
\texttt{plain} if a net without weights is required; 
\texttt{output-nonbranching} when a place may not have more than one transition in its post-set; 
\texttt{t-net} when each place may also not have more than one transition in its pre-set; 
\texttt{conflict-free} when each place is either output-nonbranching or its post-set is a subset of its pre-set; 
\texttt{$k$-bounded} if every place must never contain more than \(k\) tokens in any reachable marking; 
\texttt{safe} if the net should be 1-bounded; 
\texttt{language} if only a Petri net with the same prefix language is searched for; 
and \texttt{verbose} to print additional information about the calculated solution.
These definitions conform to those of section \ref{background.sct} and \cite{besdev-lata,best-wimmel-rostock}.
Additionally, a distributed Petri net can be requested (see below).

As an example, consider the reachability graph $lts$ shown in figure \ref{fig:example_PN}.
Let us start by just requesting any Petri net solution. This is done by running \texttt{./apt.sh synth none lts.apt}.
One possible solution is shown as $N_3$ in the same figure.
This net is similar to $N_1$ in the sense that both of them have reachability graph $lts$,
but some structural differences can be observed.
Synthesis is implemented by an algorithm \cite{bbd} involving
the solution of several systems of linear inequalities.
These solutions give rise to a large (possibly redundant) set of regions.
{}From these regions, \apt{} selects a non-redundant but still sufficiently large subset,
so that the corresponding Petri net 
also solves $lts$, provided the latter is solvable at all.
Depending on the way these inequality systems are solved,
different non-redundant sets of regions may be produced.
In some releases, \apt{} used (and incorporated) {\tt ojAlgo} (cf. \url{http://ojalgo.org/}).
Later, {\tt SMTInterpol} \cite{smt-interpol} was used.
$N_3$ is created via {\tt ojAlgo}; in other releases,
a different solution of $lts$ can and will be obtained.
The implementation is exact in the sense that if any solution exists, one will be
found. No further guarantees about the synthesised Petri net can be made.

As mentioned above, \apt{} supports the synthesis of Petri nets with special properties.
For example, suppose that we wish a synthesised net to be plain and pure.
Then we can run \texttt{./apt.sh synth plain,pure lts.apt}.
In this case, \apt{} modifies the set of inequalities handed to a solver;
the solver returns a different solution;
and \apt{}'s selection process constructs a set of non-redundant regions corresponding to the net $N_1$
shown in figure \ref{fig:example_PN}.
The same net is calculated  when \texttt{2-bounded} or just \texttt{plain} or \texttt{pure} is specified,
although none of this can be guaranteed by the implementation.

\begin{lstlisting}[float, language=shell, label={lst:synth-safe}, caption={Failure when trying to synthesise a safe Petri net from
$lts$ ($s_4$ refers to a node in figure \protect\ref{fig:example_PN})}]
$ ./apt.sh synthesize safe,verbose lts.apt
success: No
solvedEventStateSeparationProblems:
Region { init=1, 0:a:0, 0:b:0, 1:c:0, 0:d:1 }:
	separates event c at states [s4, s5, s6]
Region { init=0, 0:a:0, 0:b:1, 1:c:0, 0:d:0 }:
	separates event c at states [s0, s1, s4, s5]
[...]
failedStateSeparationProblems: []
failedEventStateSeparationProblems: {b=[s4]}
\end{lstlisting}

If we try to synthesise a safe Petri net from $lts$, we get a failure. The corresponding arguments to \apt{} and its
output are shown in listing \ref{lst:synth-safe}. This is also an example for the \texttt{verbose} option.
Each calculated region of the lts corresponds to a place in the Petri net that is being synthesised.
For example, the
first region in the above output is \texttt{\{ init=1, 0:a:0, 0:b:0, 1:c:0, 0:d:1 \}}.
This corresponds to a place with
initial marking one and from which transition \(c\) consumes a token while \(d\) produces a token each time it fires.
Also, this place disables the transition \(c\) in states \(s_4\), \(s_5\) and \(s_6\), as indicated in the output shown in listing \ref{lst:synth-safe}.
Five such regions are found, but synthesising
still fails, because no region can be calculated which disables event \(b\) in state \(s_4\)
(cf. figure \ref{fig:example_PN}). In the jargon, ``$b$ cannot be separated safely at $s_4$''.\footnote{Note that
\apt{}'s output is nevertheless correct. Every Petri net solution must have some place \(p\) which prevents \(b\) in the
marking that corresponds to \(s_4\). Since the sequence \(db\) is fireable in \(s_4\), transition \(d\) must produce
enough tokens on \(p\) to enable \(b\). Also \(ab\) is fireable, so transition \(a\) produces tokens on \(p\) as well.
Finally, the firing sequence \(ad\) is also enabled in \(s_4\). By the above reasoning, both \(a\) and \(d\) produce at
least one token on \(p\), so after \(ad\) that place must be marked with at least two token. Thus, no safe Petri net
solution exists.}

\begin{lstlisting}[float, language=APT, label={lst:synth-locations}, xleftmargin=-0.1cm,
caption={Adding locations to the reachability graph
from listing \ref{lst:example_LTS}. Only the changes are shown.}]
[...] .labels
      a[location="A"] b[location="B"] c[location="A"] d[location="A"]
[...]
\end{lstlisting}

The \texttt{synthesize} module also supports the specification of locations for transitions. If two transitions have different locations, they
must have disjoint pre-sets \cite{bd-psi11}. In both Petri nets which were synthesised so far, transitions
\(b\) and \(c\) always had a common place in their pre-sets. Next, we will look for a Petri net where \(b\)'s preset is
disjoint from the presets of all other transitions. Listing \ref{lst:synth-locations} shows how to specify this
in the \apt{} file format.
If an lts contains locations, the \texttt{synthesize} module will always honour them. No special command line option to
enable this is required. When synthesising a Petri net from the modified input file, the net $N_2$ shown in figure
\ref{fig:example_PN} is generated. It can be seen that the pre-sets of all transitions are disjoint in that net, even though the
input file only required that transition \(b\) has no common place in its pre-set with the other transitions.
In general,
specifying different locations for all transitions is tantamount to requiring an ON output net.

\begin{lstlisting}[float, language=shell, label={lst:word}, caption={Example of \texttt{word\_synthesize} in order to synthesise \(w=abbaac\).}]
$ ./apt.sh word_synthesize none a,b,b,a,a,c
success: No
separationFailurePoints: a, b, [a] b, a, a, c
\end{lstlisting}

\apt{} also provides word synthesis.
For a given word $w$, a Petri net with injective labelling is produced such that $w$ and its prefixes are the only
enabled firing sequences. Given a word \(w=a_1a_2\dots a_n\), this module internally creates an lts
\((S,\rightarrow,T,s_0)\) with \(n+1\) states \(S=\lbrace s_0,s_1,\dots s_n\rbrace\), transitions \(T=\lbrace
a_1,a_2,\dots a_n\rbrace\), and transition relation \(\rightarrow=\lbrace (s_{i-1}, a_i, s_i) \mid
i\in\lbrace 1,2,\dots,n\rbrace\rbrace\).
Listing \ref{lst:word} shows an application. 
In the first line, \apt{} is asked to synthesise the word \(abbaac\) (specified as a comma-separated list).
The set of transitions is implicitly assumed to be \(T=\lbrace a,b,c\rbrace\).
No requirements are specified for the synthesised Petri net, and still, a failure occurs.
The output shows that after the subword \(ab\), the transition
\(a\) is enabled, even though the input requires the transition \(b\)
to be the only enabled transition.\footnote{This
result is correct since $a$ cannot be separated at state $s_2$.
That is,
any injectively labelled Petri net in which the word \(abbaac\) and all of its prefixes are fireable,
must also have a firing sequence $aba$.}

\subsection{Some algorithmic background}

By courtesy of the authors of \cite{bbd}, the authors were fortunate to be able to use an advance draft of \cite{bbd}
when implementing the \texttt{synthesize} module.
Nevertheless, for the purpose of creating solutions with special properties, it was necessary to extend the theory somewhat.
Some of these amendments are described (very briefly) in the following.
\apt{} contains a generic implementation that can handle all of the supported properties,
and for some special cases, \apt{} contains faster algorithms.

Formally, a region of an lts $(S,\to,T,s_0)$ is
a triple $(\mathbb{R},\mathbb{B},\mathbb{F})\in(S\to\nsymbol,T\to\nsymbol,T\to\nsymbol)$
such that for all $s[t\rangle s'$ with $s\in[s_0\rangle$,
$\mathbb{R}(s)\geq\mathbb{B}(t)$ and $\mathbb{R}(s')=\mathbb{R}(s)-\mathbb{B}(t)+\mathbb{F}(t)$.
Essentially, $\mathbb{B}$ and $\mathbb{F}$ assign {\em backward and forward weights} to transitions $t$ of an lts,
so that these weights can serve as connecting arc weights between $t$ and a place of a Petri net,
and $\mathbb{R}$ assigns a token count in each marking to that place.
The derived function $\mathbb{E}\colon T\to\zsymbol$ defined by $\mathbb{E}(t)=\mathbb{F}(t)-\mathbb{B}(t)$
is called the {\em effect} of a transition $t$.
Because the effect is zero around cycles of the lts, the functions $\mathbb{B}$ and $\mathbb{F}$ necessarily satisfy
\(\sum_{t\in T}\Psi(t)\cdot\mathbb{B}(t)=\sum_{t\in T}\Psi(t)\cdot\mathbb{F}(t)\)
for every cyclic Parikh vector \(\Psi\) in the lts.
A region is called {\em pure} if it satisfies \(\forall t\in T\colon\mathbb{B}(t)=0\lor\mathbb{F}(t)=0\).


For synthesising a Petri net from an lts, regions solving \emph{separation problems} have to be found. There are two kinds of such
problems. For each state \(s\) in which transition \(t\) is not enabled, there is an \emph{event/state separation problem}
\(\mathbb{R}(s)<\mathbb{B}(t)\) that corresponds to a place preventing the transition \(t\).
For each pair of states \(\lbrace s,s'\rbrace\) with \(s\neq s'\)
there is a \emph{state separation problem} \(\mathbb{R}(s)\neq\mathbb{R}(s')\)
so that these states are represented by different markings.
The task at hand is to find, for any given separation problem, a region that solves it.
A set \(R\) of regions is feasible for synthesising a Petri net if each
separation problem is solved by at least one of its regions. In this case the Petri net described by \(R\) solves the lts.
However, since special properties might be requested from the calculated Petri net, only regions which do not contradict these
properties should be used.
Some algorithms optimise the search for feasible regions but do not allow special properties to be guaranteed.
Others are less efficient in general but more flexible in terms of the result.
\apt{} chooses an appropriate algorithm, which may depend on the result specification, as follows.

{\bf Petri net synthesis with additional properties.}
\apt{} comes with a general algorithm supporting all properties.
For this, first a region basis is calculated from the cycles of the transition system. This basis has the property that all pure
regions are a linear combination of its elements. An inequality system is used for finding such a combination.
For solving a specific separation problem, the initial marking \(\mathbb{R}(s_0)\) and the backward and forward weights
\(\mathbb{B}(t)\) and \(\mathbb{F}(t)\) for every transition \(t\) are variables.
With these, we explicitly require for any state \(s'\in S\) and enabled transition \(t\in T\) that the region does
not block $t$.
This can be expressed via \(s'[t\rangle\implies\mathbb{R}(s')=\mathbb{R}(s_0)+\mathbb{E}(\Psi_{s'})\geq \mathbb{B}(t)\),
where $\Psi_{s'}$ is the Parikh vector of the path from $s_0$ to $s'$ in some fixed spanning tree.
Then, any solution of the system describes a valid region of the lts under consideration.
For separating states \(s\) and \(s'\), an additional inequality \(\mathbb{R}(s)\neq\mathbb{R}(s')\) is required. Since for each
place of a bounded Petri net, a complementary place can be added so that the token sum of the two places stays constant,
this inequality can be softened to \(\mathbb{R}(s)<\mathbb{R}(s')\).
For separating transition \(t\) from state \(s\), either
$\mathbb{R}(s)-\mathbb{B}(t) < 0$ or $\mathbb{R}(s)+\mathbb{E}(t) < 0$ is used,
depending on whether the resulting Petri net should be impure or pure.

Additional inequalities are added to guarantee the requested properties.
When locations are specified, only transitions on the same location as \(t\) may have \(\mathbb{B}(t)>0\),
i.e., may consume token from this place in the final Petri net. For all other transitions \(t'\) the equation \(\mathbb{B}(t')=0\)
makes sure that no conflict between locations occurs.
Calculating output-nonbranching solutions makes use of this by internally assigning a unique location to each transition.
If the user asks for a plain solution, the algorithm adds \(\mathbb{B}(t)\leq 1\) and \(\mathbb{F}(t)\leq 1\) for every
transition \(t\in T\) to the inequality system.
T-nets are found by requiring a plain solution where additionally the sum of all forward weights is at most one,
\(1\geq\sum_{t\in T}\mathbb{F}(t)\), and the same for backward weights.
If a conflict-free net should be synthesised, plainness is additionally required and the implementation first searches an
output-nonbranching region and, if this fails, the corresponding inequalities are replaced with \(\mathbb{E}(t)\geq 0\)
for all transitions \(t\). This ensures that the preset contains the postset of the place that corresponds to the
calculated region.
Finally, calculating a \(k\)-bounded Petri net requires adding an inequality \(k\geq\mathbb{R}(s)\) for each state \(s\).
Because this is, so far, the only property that requires adding an inequality for each state, it is the most expensive one.

{\bf Speeding up general Petri net synthesis.}
If the \texttt{synthesize} module is invoked just with result specification \texttt{none}, and no locations are specified,
synthesis can be made more efficient.
The approach for event/state separation is to calculate a region where \(\mathbb{R}(s)\) is smaller than \(\mathbb{R}(s')\) for any
state \(s'\) in which transition \(t\) is enabled. Then both \(\mathbb{B}(t)\) and \(\mathbb{F}(t)\) can be increased
by the same amount (possibly introducing side conditions) so that the transition becomes separated.
To find such a region, the system \(\forall s'\in S\colon s'[t\rangle\implies\mathbb{E}(\Psi_s-\Psi_{s'})<0\) has to be
solved, where the weights of the region basis are the unknowns (that is, a much smaller system
has to be solved).
For state separation, the regions from the region basis can be tested and used. This is because if the regions from the
basis do not separate \(s\) and \(s'\), then no linear combination of the basis elements will
either.\footnote{Note: This algorithm and the previous one (without additional properties beside pure)
are described in detail in \cite{bbd}.}

{\bf Pure and pure\&plain Petri net synthesis.}
Suppose that the result request is \texttt{pure}, or \texttt{pure,plain} (read conjunctively),
and that again, no locations are supported.
For solving state separation, if only a pure solution is requested, the previous approach can be used, because all
elements of the region basis calculated there are pure regions.
For separating transition \(t\) from state \(s\), by definition, a region satisfying \(\mathbb{R}(s)<\mathbb{B}(t)\) is needed.
Since \(\mathbb{R}\) can be calculated based on the value
\(\mathbb{R}(s_0)\) for the initial state and the Parikh vector \(\Psi_s\), this is equivalent to
\(\mathbb{R}(s_0)+\mathbb{E}(\Psi_s)-\mathbb{B}(t)<0\).
After more simplifications, we see that we have to solve \(\forall s'\in S\colon
\mathbb{E}(\Psi_s-\Psi_{s'}+\mathbf{1}_t)<0\) where $\mathbf{1}_t$ is the $t$-unit vector.
As before, the resulting region has to be a linear combination of the region basis.
If a plain Petri net should be calculated, additional constraints are added that ensure that \(-1\leq \mathbb{E}(t)\leq 1\) for
all transitions \(t\) i.e., that the forward and backward weights are either one or zero.

{\bf Synthesising marked graph Petri nets.}
The reachability graphs of marked graphs are characterised and a special synthesis algorithm is presented in
\cite{besdev-lata}. This algorithm calculates a Petri net solution directly, based on structural properties of the lts,
and is implemented in \apt{}. The details will not be repeated in
the current paper. Suffice it to say that \apt{}'s \texttt{synthesize} module automatically checks the required structural
preconditions on the lts and uses the improved algorithm if it is applicable. This algorithm supports any
combination of the properties \texttt{pure}, \texttt{plain}, and \texttt{t-net}, and any location specification.

{\bf Synthesis up to language equivalence.}
If a Petri net with the same prefix language as the input lts is needed, a so-called limited unfolding of the lts \cite{bbd} is
calculated. This unfolding is synthesised as usual, but without enforcing state separation.

{\bf Heuristically minimizing the number of places.}
A feasible set of regions
could stay feasible if some regions are removed from it. This can occur because regions calculated for a specific
separation problem could additionally solve other separation problems. Thus, it makes sense to remove unnecessary
regions from the set of calculated regions.
For this, all event/state separation and state separation problems are evaluated again in the regions found. If such a
problem is solved by just a single region, that region cannot be removed from the feasible set of regions. This region
is called a \emph{required} region. Any separation problem which is solved by a required regions can be discarded. For the
remaining problems which are solved by multiple non-required regions, any of these regions could be picked arbitrarily. In
practice this heuristic produces Petri nets with an acceptably low number of places.\footnote{This heuristic introduces nondeterminism.
Alternatively, some total ordering could be imposed on regions to break ties.}

\subsection{Benchmarks}
\label{benchmark.sct}

The performance of \apt{}\footnote{The latest development version was used. It can be identified by git commit id
{\tt 14651f7280db255d1539} in \cite{apt}.} for Petri net synthesis was compared with {\tt synet} 2.0b \cite{synet},
{\tt petrify} 4.2 \cite{petrify} and {\tt GENET} \cite{genet} on a system running Fedora 21 with an Intel\textregistered{}
Core\texttrademark{} i7-4790 CPU clocked at 3.6 GHz and with 32 GiB of memory.
The {\tt synet} tool can synthesise distributable bounded Petri nets.
For {\tt petrify}, the user can choose between some
properties, for example pure, free choice and unique choice.
However, {\tt petrify} only creates safe Petri nets and employs
transition splitting to ensure that a solution exists.
This means that the resolution Petri nets might not be injectively labelled.
With {\tt GENET}, the result will only be bisimilar to the input. Also, this tool requires {\it a priori} knowledge about the
maximum number of token on any place, and it resorts to transition splitting to produce solutions.
Given these differences, it can be expected that {\tt petrify} and {\tt GENET} perform
better on safe nets and worse on transition systems which have no safe solution.

\begin{table}
	\centering
	\begin{tabular}{r|rrrrr|rrrrr}
		    & \multicolumn{5}{|c|}{bit net synthesis} & \multicolumn{5}{|c}{philosophers' net synthesis}\\\hline
		$n$ & APT & APTp     & synet & petrify & GENET & APT & APTp     & synet & petrify & GENET \\\hline
		8  &   0.60 &   0.86 & 138.49 &   0.13 &   0.05 &   0.55 &   0.49 &   0.06 &   0.01 &   0.01 \\
		10 &   1.56 &   2.32 & ---    &   1.25 &   0.31 &   0.50 &   0.60 &  10.08 &   0.05 &   0.03 \\
		12 &   5.71 &   6.31 & ---    &  17.73 &   2.28 &   0.79 &   1.05 & ---    &   0.25 &   0.09 \\
		14 &  24.69 &  30.48 & ---    & 403.67 &  16.10 &   1.72 &   2.42 & ---    &   0.91 &   0.33 \\
		16 & 183.76 & 212.23 & crash  & ---    & 132.13 &   4.49 &   5.21 & ---    &   4.11 &   1.31 \\
		18 & ---    & ---    & crash  & OOM    & ---    &   9.17 &  13.13 & ---    &  21.84 &   4.83 \\
		20 &        &        &        &        &        &  26.76 &  41.96 & ---    & 171.10 &  19.88 \\
		22 &        &        &        &        &        &  98.57 & 146.42 & crash  & ---    & 123.05 \\
	\end{tabular}
	\caption{Time in seconds for synthesising a Petri net. APTp means \apt{} with the {\tt pure} parameter.
		Dashes indicate that the 10 minutes time limit was exceeded.
		For large inputs, {\tt synet} crashed with a stack overflow and {\tt petrify} exited with a memory allocation error.}
	\label{tbl:bench-bitnet-bistate-results}
\end{table}

Three of \apt{}'s Petri nets generators were used.
The \texttt{bitnet\_generator} module creates a net where
\(n\) bits can be flipped between two states, creating $2^n$ states in total.
The \texttt{bistate\_philnet\_generator} model Dijkstra's philosophers problem \cite{dij-phil} for \(n\) philosophers
such that each philosopher grabs both forks in a single step and puts them back simultaneously as well.
The \texttt{cycle\_generator} creates a cycle consisting of \(n\) transitions and \(n\) places where \(k\) tokens are
moved from one place to the next in a cyclic way.
All these generators produce plain and pure nets. The first two generators and cycles with $k=1$ are additionally safe.
In this case, all contesting tools can
correctly synthesise nets from the reachability graph of the generated nets, although {\tt GENET} might produce a net
which only exhibits bisimilar behaviour. For $k>1$, transition splitting will be done by {\tt petrify} and {\tt GENET}.

{\tt petrify} was used with argument \texttt{-dead}, so that it does not complain about
deadlocks. \apt{} was measured for general synthesis and for
pure synthesis. In contrast to {\tt petrify}, which produced similar run times in these two cases, this makes a
difference for \apt{}. {\tt synet} was only benchmarked with parameter \texttt{-r}, since it
performed consistently worse without this argument. {\tt GENET} was used without any arguments.
Measurements were made by generating the reachability graph of the net that the Petri net generator produced, converting
the net into the input format of each tool with \apt{} and then measuring the wall clock time needed by each tool to synthesise a Petri net
from this graph. The time for synthesis was limited to 10 minutes via the \texttt{ulimit -t} unix command.
For each input, three measurements were taken, out of which the minimal values are depicted in
Tables \ref{tbl:bench-bitnet-bistate-results} to \ref{tbl:bench-cycle-multiple-tokens-results}.

The result for the class of bit nets are shown
in the left part of table \ref{tbl:bench-bitnet-bistate-results}.
It can be seen that with 18
bits, none of the tools managed to find a solution within the 10 minutes time limit. This table also shows that \apt{} has a
relatively high start-up cost, causing it to require more time for small inputs.
Also, \apt{} only slows down moderately
if a pure solution is requested. Surprisingly, {\tt synet} crashes with a stack overflow error if the input becomes too
large and {\tt petrify} runs out of memory for the reachability graph of a 17 bit (not shown) or 18 bit net. Its peak memory usage is about 1
GiB, so the system's physical memory is not exhausted. In this benchmark, {\tt GENET} is a bit faster than \apt{}.

Table \ref{tbl:bench-bitnet-bistate-results} also contains the results for the philosophers' nets in its right part.
Here \apt{} outperforms {\tt GENET}, but only for the largest inputs.
Up to $n=20$, {\tt GENET} is consistently faster.
When requesting a pure
solution, \apt{} becomes slower than {\tt GENET} searching for any solution at all. When compared to to {\tt petrify},
similar behaviour can be seen, although here the crossing point is at $n=17$.
In this experiment \apt{} is still faster than {\tt GENET} if a pure solution is requested.

\begin{table}
	\centering
	\begin{tabular}{r|rrrrrr}
		$n$ & APT & APTp & synet & petrify & GENET \\\hline
		100 &  0.44 &  0.45 &   1.12 & 0.02 &  0.28 \\
		180 &  1.58 &  1.58 &   8.83 & 0.05 &  1.81 \\
		260 &  5.44 &  5.45 &  35.99 & 0.10 &  6.34 \\
		340 & 16.45 & 16.05 & 102.52 & 0.15 & 17.45 \\
		420 & 40.55 & 40.90 & 234.59 & 0.23 & 32.99 \\
		500 & 83.15 & 83.53 & 475.50 & 0.32 & 62.39
	\end{tabular}
	\caption{Cycle synthesis run times with cycles of size $n$ and $k=1$ token.}
	\label{tbl:bench-cycle-results}
\end{table}

The times for the cycle nets with a single token are shown in table \ref{tbl:bench-cycle-results}. Compared to the other examples,
these nets show no concurrent behaviour and are about as large as their reachability graphs. In this benchmark, \apt{}
uses its implementation of the marked graph synthesis from \cite{besdev-lata}.
Still, {\tt petrify}, for reasons not known to the authors, almost needs no time at all.

\begin{table}[htb]
	\centering
	\begin{tabular}{r|rrrrr||r|rrrrrrr}
		\multicolumn{6}{c||}{size $n$ varying, $k=5$ tokens fixed} & \multicolumn{6}{c}{size $n=5$ fixes, $k$ tokens varying}\\\hline
		$n$ & APT & APTp & synet & petrify & GENET & $k$ & APT & APTp & synet & petrify & GENET \\\hline
		 5 & 0.19 & 0.19 & 0.00  & 10.08  & 136.52 &  5 & 0.19 & 0.19 &   0.00 & 10.08  & 136.52 \\
		10 & 0.49 & 0.51 & ---   & ---    & 468.38 & 10 & 0.37 & 0.30 &   0.16 & ---    & 292.74 \\
		15 & 1.35 & 1.39 & ---   & ---    & ---    & 15 & 0.61 & 0.72 &   3.19 & ---    & --- \\
		20 & 4.83 & 4.58 & ---   & ---    & ---    & 20 & 2.00 & 1.14 &  16.47 & ---    & --- \\
\multicolumn{1}{r}{}&     &      &       &        &        & 25 & 2.42 & 2.09 &  93.22 & ---    & --- \\
\multicolumn{1}{r}{}&     &      &       &        &        & 30 & 4.16 & 3.81 & 190.81 & ---    & ---
	\end{tabular}
	\caption{Cycle synthesis run times with cycles of size $n$ and $k$ tokens. Left part varies size of cycle, right
		part varies number of token.}
	\label{tbl:bench-cycle-multiple-tokens-results}
\end{table}

When synthesising cycles with $k=5$ tokens, the cycles have to be a lot smaller. The corresponding result are shown
in the left part of table \ref{tbl:bench-cycle-multiple-tokens-results},
and it can be seen that the tools that use transition splitting need much longer.
The debug output suggests that the splitting leads to an exponential increase in the state space.
Also, {\tt synet} only manages to synthesise the smallest cycle size within the time limit. In contrast to this, \apt{}
produces results quickly, because in this case, the marked graph synthesis algorithm performs optimally. The results for cycles of
size $n=5$ with increasing numbers of tokens are similar and can be found in
the right part of the same table.
The main difference is that {\tt synet} performs a lot better when the number of tokens is increased instead of
enlarging the size of the cycle.

An experiment was done by hand for cycles of size $n=3$ with $k=100$ tokens. In this setup, \apt{} needed 0.65
seconds to find a solution, {\tt synet} finished in 0.98 seconds and \apt{} with parameter \texttt{pure} in 1.03
seconds.
{\tt GENET} ran out of memory after allocating 4 GiB in 422 seconds. After 40 minutes without any result, {\tt petrify} was aborted.
In this special case, {\tt GENET} was also measured with parameter \texttt{-k 100}, telling it to look for 100-bounded
solutions, and found one in 7.67 seconds. When the search with bounds 1 to 99 was skipped
via parameters \texttt{-k 100 -min 100}, {\tt GENET} needed only 2.20 seconds.
This confirms previous intuitions that transition splitting may lead to bad run times
(and, of course, to non-injectively labelled nets), but it also shows that {\tt GENET} is sped up if {\it a priori} knowledge is available.
Still, even for the safe case, \apt{} has comparable results and has been generalised (like {\tt synet}) to unsafe nets.

\section{Concluding remarks}
\label{concl.sct}

{\tt APT}'s algorithms are packaged in a single, portable archive called {\tt apt.jar}.
The idea is that a user can copy this file and
run it smoothly, using his or her favourite text editors, in a local Java 7 environment,
or alternatively, grab the entire {\tt apt} directory from the repository at \cite{apt} and build a local copy
of {\tt apt.jar} using {\tt ANT}.
{\tt APT}'s performance in its other modules (for example, {\tt coverability})
was tested against other tools (for example, {\tt LoLA 2.0} \cite{lola})
and seems to perform worse, but not hopelessly so.\footnote{The authors are
aware of (and have tested) a multitude of other Petri net tools.
Not all of them could be mentioned in this paper.} 
In general, the authors hope that all of \apt{}'s modules can be used sensibly in a classroom environment,
say for a course on place/transition Petri nets and finite transition systems.
They also believe that \apt's more sophisticated algorithms
can, in addition, be helpful to researchers in the corresponding areas.

In future, we wish to explore whether code written, say, in {\tt C++}
could be incorporated into {\tt APT} more tightly than just
by means of exchanging text files for nets and transition systems.
Also, graphical extensions will be explored cautiously (cf. \cite{hillit}).
However, before imposing a more powerful user interface onto {\tt APT},
we would like to explore intelligent -- possibly interactive -- extensions.
For instance, consider the algorithm testing the strong small cycle property.
If no prior assumptions hold, it is nontrivial and, in general, rather time-consuming.
However, suppose that the preconditions of the result mentioned at the end of section \ref{background.sct} have
already been tested and are known to hold for the given lts.
Then we know that the weak small cycle property also holds,
and testing the strong one is much easier.
(The same principle -- using theory to algorithmic advantage -- is behind \apt's fast marked graph synthesis.)
It is also planned to extend word synthesis to the prefix languages of regular languages.
This is pretty straightforward, since it is
well-known how to construct an lts from a regular expression.
Other extensions could consist of parallelising some of the algorithms.
Dennis Borde, one of the {\tt APT} students, already succeeded in parallelising part of the coverability graph
generation algorithm by exploiting the power of a graphics card processor
running concurrently with the main processor.

{\bf Acknowledgements:} 
The authors would like to thank the reviewers for helpful comments.

\appendix

\section{Labelled transition systems and Petri nets}
\label{background.sct}

An lts (labelled transition system with initial state)
is a tuple $(S,\to,T,s_0)$, where
$S$ is a set of {\em states};
$T$ is a set of {\em labels} with $S\cap T=\es$;
$\to\subseteq(S\times T\times S)$ is the {\em transition relation};
and $s_0\in S$ is an {\em initial state}.
A label $t$ is {\em enabled} in a state $s$,
denoted by $s[t\rangle$, if there is some state $s'$ such that $(s,t,s')\in\to$.
$s[t\rangle s'$ ($s[\tau\rangle s'$)
means that $s'$ is {\em reachable} from $s$
through the execution of $t$ (resp., of $\tau\in T^*$). 
By $[s\rangle$, we denote the set of states reachable from $s$.
For $\sigma\in T^*$, the {\em Parikh vector}
$\Parikh(\sigma)$ is a $T$-vector 
where $\Parikh(\sigma)(t)$ denotes
the number of occurrences of $t$ in $\sigma$.
$s[\sigma\rangle s'$ is called a {\em cycle} 
if $s=s'$, and $\Parikh(\sigma)$ is called {\em cyclic} in this case. 
A nontrivial cycle $s[\sigma\rangle s$ around a reachable state $s\in[s_0\rangle$
is called {\em small} if
there is no nontrivial cycle $s'[\sigma'\rangle s'$ with $s'\in[s_0\rangle$ and $\Parikh(\sigma')\lneqq\Parikh(\sigma)$.

Two lts $(S_1,\to_1,T,s_{01})$ and $(S_2,\to_2,T,s_{02})$ over the same set of labels $T$ are
{\em language-equivalent} if their initially enabled sequences coincide,
i.e., if $\forall\sigma\in T^*\colon s_{01}[\sigma\rangle\iff s_{02}[\sigma\rangle$,
{\em isomorphic} if there is a bijection $\zeta\colon S_1\to S_2$ with $\zeta(s_{01})=s_{02}$ and
$(s,t,s')\in\to_1\iff(\zeta(s),t,\zeta(s'))\in\to_2$, for all $s,s'\in S_1$;
and {\em bisimilar} if there is a relation
$\beta\subseteq S_1\times S_2$ with $(s_{01},s_{02})\in\beta$
and whenever $(r_1,r_2)\in\beta$ and $(r_1,t,s_1)\in\to_1$, then $\exists s_2\in S_2\colon(r_2,t,s_2)\in\to_2$
(and {\it vice versa}).

A labelled transition system $(S,\to,T,s_0)$ is called
{\em finite} if $S$ and $T$ (hence also~$\to$) are finite sets;
{\em deterministic} if for any reachable state $s$ and label $a$, $s[a\rangle s'$ and $s[a\rangle s''$ imply $s'=s''$;
{\em totally reachable} if $S=[s_0\rangle$ and $\forall t\in T\exists s\in[s_0\rangle\colon s[t\rangle$;
{\em reversible} if $\forall s\in[s_0\rangle\colon s_0\in[s\rangle$;
{\em persistent} if for all reachable states $s$ and labels $t,u$,
if $s[t\rangle$ and $s[u\rangle$ with $t\neq u$,
then there is some state $r\in S$ such that both $s[tu\rangle r$ and $s[ut\rangle r$.
It has the {\em weak small cycle property} if there is a finite set of mutually transition-disjoint Parikh vectors
such that every small cycle has a Parikh vector in this set, and the {\em (strong) small cycle property}
if every small cycle has the same Parikh vector.


A (finite, initially marked, place-transition, arc-weighted)
Petri net is a tuple $(P,T,F,M_0)$ such that
$P$ is a finite set of {\em places},
$T$ is a finite set of {\em transitions}, with $P\cap T=\es$,
$F$ is a {\em flow} function $F\colon((P\times T)\cup(T\times P))\to\nsymbol$,
$M_0$ is the {\em initial marking},
where a {\em marking} is a mapping $M\colon P\to\nsymbol$, indicating the number of {\em tokens} in each place.
A transition $t\in T$ is {\em enabled by} a marking $M$,
denoted by $M[t\rangle$, if for all places $p\in P$, $M(p)\geq F(p,t)$.
If $t$ is enabled at $M$, then $t$ can {\em occur} (or {\em fire}) in $M$,
leading to the marking $M'$ defined by $M'(p)=M(p)-F(p,t)+F(t,p)$
(notation: $M[t\rangle M'$).
The {\em reachability graph of $N$}, with initial marking $M_0$,
is the labelled transition system with the set of vertices $[M_0\rangle$ (i.e., the states which are reachable from $M_0$)
and set of edges $\{(M,t,M')\mid M,M'\in[M_0\rangle\land M[t\rangle M'\}$.
If an lts $TS$ is isomorphic to the reachability graph of a Petri net $N$, then we will
also say that $N$ {\em solves} $TS$.
If $k$ is a natural number and $M$ a marking, then $k{\cdot}M$
denotes the marking with $(k{\cdot}M)(p)=k{\cdot}M(p)$ for every place $p$.

For a place $p$ of a Petri net $N=(P,T,F,M_0)$, let ${}^\dt p=\{t\in T\mid F(t,p)>0\}$
its pre-places, and $p^\dt=\{t\in T\mid F(p,t)>0\}$ its post-places.
$N$ is called {\em (strongly/weakly) connected} if it is strongly/weakly connected as a graph;
{\em plain} if $cod(F)\subseteq\{0,1\}$;
{\em pure} or {\em side-condition free} if $p^\dt\cap{}^\dt p=\es$ for all places $p\in P$;
ON ({\em place-output-nonbranching}) if $|p^\dt|\leq1$ for all places $p\in P$;
CF ({\em conflict-free}) if it is plain and $\forall p\in P\colon|p^\dt|>1\impl p^\dt\subseteq{}^\dt p$;
BCF ({\em behaviourally conflict-free})
if it is plain and for any two transitions $t,t'\in T$ with $t\neq t'$ and
for every $M\in[M_0\rangle$,
if $M[t\rangle$ and $M[t'\rangle$ then ${}^\dt t\cap{}^\dt t'=\es$;
BiCF ({\em binary-conflict-free})
if it is plain and for any two transitions $t,t'\in T$ with $t\neq t'$ and
for every $M\in[M_0\rangle$,
if $M[t\rangle$ and $M[t'\rangle$ then $\forall p\in P\colon M(p)\geq F(p,t){+}F(p,t')$;
a {\em marked graph} ({\em T-net}) if it is plain and $|p^\dt|=1$ and $|{}^\dt p|=1$
(resp., $|p^\dt|\leq1$ and $|{}^\dt p|\leq1$) for all places $p\in P$; 
{\em weakly live} if
$\forall t\in T\exists M\in[M_0\rangle\colon M[t\rangle$
(i.e., there are no unfireable transitions);
{\em $k$-bounded} for some fixed $k\in\nsymbol$, if
$\forall M\in[M_0\rangle\forall p\in P\colon M(p)\leq k$
(i.e., the number of tokens on any place never exceeds $k$);
{\em bounded} if
$\exists k\in\nsymbol\colon N\text{ is $k$-bounded}$;
{\em persistent} ({\em reversible}) if so is its reachability graph.
For a number $k\in\nsymbol$,
a net with marking $k{\cdot}M$ is called {\em strongly separable from $k{\cdot}M$}
if every firing sequence starting at $k{\cdot}M$ belongs to the shuffle
product of $k$ firing sequences starting at $M$,
and {\em weakly separable from $k{\cdot}M$} if the Parikh vector of every firing sequence
starting at $k{\cdot}M$ is the sum of the Parikh vectors of $k$ firing
sequences starting at $M$.

A {\em labelled Petri net} has, in addition, a labelling function $h\colon T\to\Sigma$
where $\Sigma$ is some set of transition labels. This induces a double labelling
of the arcs of corresponding reachability graph: first, with transitions of $T$, and then,
with labels from $\Sigma$.
In case a net is labelled, the definitions of language-equivalence, isomorphism and bisimulation
are the same as previously, except that they are taken with respect to $\Sigma$.
If a net is unlabelled, $\Sigma=T$ is assumed implicitly (and explicitly in {\tt APT}).

The interest of the small cycle property arises from the following result \cite{bd-acta}:
{\em The reachability graph of a bounded, weakly live, reversible, persistent Petri net $N$ is finite
and satisfies the weak small cycle property.}
If one requires connectedness and replaces ``persistent'' by ``ON'',
then the strong small cycle property can be deduced.
This suggests a close relationship between persistent lts having the small cycle property and ON Petri nets,
motivating a question which was raised in \cite{bd-psi11}:
{\em If an lts is Petri net solvable, reversible, persistent, and has the small cycle property,
does there always exist an ON Petri net generating it?}
The answer is negative, even if the critical Parikh vector is $1$ and further conditions are imposed \cite{besdev-lata}.
The search for a counterexample turned out to be tedious,
and was, in fact, one of the reasons for initiating \apt.
Another reason was the desire for tool support in examining further open questions,
such as the following one from \cite{bd-fi}:
{\em Is the reachability graph of a plain, pure, bounded, reversible, persistent net
with an initial marking $K{\cdot}M$ with $K\geq2$ always isomorphic
to the reachability graph of some marked graph?}

\end{document}